\documentclass[useAMS,usenatbib]{mn2e}
\usepackage{epsfig}
\usepackage{amsmath}
\usepackage{rotating}

\usepackage{multirow}
\usepackage{amssymb}
\usepackage{graphicx}
\usepackage{graphics}
\usepackage{rotating}
\usepackage{lscape}

\newcommand{\be}{\begin{equation}}
\newcommand{\beq}{\begin{equation}}
\newcommand{\ba}{\begin{eqnarray}}
\newcommand{\ee}{\end{equation}}
\newcommand{\eeq}{\end{equation}}
\newcommand{\ea}{\end{eqnarray}}

\def\lsim{~\rlap{$<$}{\lower 1.0ex\hbox{$\sim$}}}

\def\gsim{~\rlap{$>$}{\lower 1.0ex\hbox{$\sim$}}}

\title[Limits on far-IR emission from CS dust...]
  {Herschel limits on far-infrared emission 
    from circumstellar dust around three nearby Type Ia supernovae}
\author[J. Johansson et al.]
  {Joel Johansson,$^1$
  Rahman Amanullah,$^1$ 
   Ariel Goobar,$^1$ 	
  \\$^1$Oskar Klein Centre, Stockholm University, SE 106 91 Stockholm, Sweden
  }
\date{Accepted 2013 January 9. Received 2013 January 9; in original form 2012 September 18}

\pagerange{\pageref{firstpage}--\pageref{lastpage}} \pubyear{2012}

\def\LaTeX{L\kern-.36em\raise.3ex\hbox{a}\kern-.15em
    T\kern-.1667em\lower.7ex\hbox{E}\kern-.125emX}

\begin{document}

\label{firstpage}

\maketitle

\begin{abstract}
  We report upper limits on dust emission at far-infrared (IR) wavelengths from
  three nearby Type Ia supernovae: SNe 2011by, 2011fe and 2012cg. 
  Observations were carried out at $70\,\mu$m and $160\,\mu$m with the {\it
    Photodetector Array Camera and Spectrometer} (PACS) on board the
  {\it Herschel Space Observatory}. 
  None of the supernovae were detected in the far-IR, allowing us to place upper 
  limits on the amount of pre-existing dust in the circumstellar environment. 
  Due to its proximity, SN 2011fe provides the tightest constraints,
  $M_{\rm dust} \lsim 7 \times 10^{-3}\,\mathrm{M}_{\odot}$ at a $3\sigma$-level 
  for dust temperatures $T_{\rm dust} \sim 500\, \mathrm{K}$ 
  assuming silicate or graphite dust grains of size $a = 0.1\, \mu$m.
  For SNe 2011by and 2012cg the corresponding upper limits are less stringent, with
  $M_{\rm dust} \lsim 10^{-1}\,\mathrm{M}_{\odot}$ for the same assumptions.
\end{abstract}

\begin{keywords}
circumstellar matter -- supernovae: general -- supernovae: individual: SN 2011by -- supernovae: individual: SN 2011fe -- supernovae: individual: SN 2012cg -- dust, extinction.
\end{keywords}
\section{Introduction}
The use of Type Ia supernovae (SNe Ia) as distance indicators remains
essential for the study of the expansion history of the Universe and
for explorations of the nature of dark energy \citep{goobar2011}.
However, a lack of understanding of the progenitor systems and the
requirement for empirically derived colour-brightness corrections
represent severe limitations for precision cosmology.  Information
about the progenitor systems of SNe Ia can be obtained by searching
for evidence of circumstellar material (CSM) associated with mass-loss
prior to the explosion.  In the single-degenerate model, a white dwarf
(WD) accretes mass from its hydrogen-rich companion star until it
reaches a mass close to the Chandrasekhar mass, at which point carbon
ignites, triggering a thermonuclear explosion.  In the
double-degenerate model, a supernova results from the merger of two
WDs.  Thus, the detection of CSM arising from the transfer of matter
to the WD by its non-degenerate binary companion would be a direct
confirmation of the single-degenerate scenario.  Dust may also be
created in the circumstellar environment before the explosion, which
would have important implications for observed colours of SNe Ia.
This second scenario is the focus of this paper.
The existence of CSM around nearby SNe Ia has been suggested by
studies of sodium absorption lines \citep[e.g. SNe 1999cl, 2006X and
2007le, ][]{patat2007,blondin2009,simon2009,sternberg2011}.
High-resolution spectra reveal the presence of time-variable and
blueshifted Na I D features, possibly originating from CSM within the
progenitor system.  Studies of large samples of SNe Ia
\citep{sternberg2011} find that half of all SNe Ia with detectable Na
I D absorption at the host-galaxy redshift have Na I D line profiles
with significant blueshifted absorption relative to the strongest
absorption component, which indicates that a large fraction of SN Ia
progenitor systems have strong outflows. \citet{foley2012} also find
that SNe Ia with blueshifted circumstellar/interstellar absorption
systematically exhibit higher ejecta velocities and redder colours at
maximum brightness relative to the rest of the SN Ia population.

Non-standard reddening has been noted in studies of individual and
large samples of SNe Ia. For example, the colour excess indices of SN 2006X
were studied in \citet{folatelli2010}, showing that the reddening is
incompatible with the average extinction law of the Milky Way. Their
findings augmented the large body of evidence indicating that the
reddening of many SNe Ia show a steeper wavelength dependence ($R_{V}
< 3.1$) than that which is typically observed for stars in our
Galaxy. Previously, \citet{nobili2008} derived $R_{V}=1.75 \pm 0.27$
from a statistical study of 80 low redshift SNe Ia. Similarly, when
the colour-brightness relation is fitted jointly with cosmological
parameters in the SNe Ia Hubble diagram, using a wide range of SNe Ia
redshifts, low values of $R_{V}$ are obtained \citep[see e.g.][for a recent
compilation]{suzuki2012}.

\citet{wang2005} and \citet{goobar2008} showed that multiple
scattering on circumstellar (CS) dust could potentially help to
explain the low values of $R_{V} \sim 1.5 - 2.5$ observed in the sight
lines of nearby SNe Ia.  \citet{rahman2011} simulated the impact of
thin CS dust shells located at radii $r_{\rm d} \sim 10^{16} -
10^{19}\,\mathrm{cm}$ ($\sim 0.003 - 3\,\mathrm{pc}$) from the SN,
containing masses $M_{\rm dust} \sim 10^{-4}\,\mathrm{M}_{\odot}$, and
find that this scenario would also perturb the optical lightcurve
shapes and introduce ``intrinsic'' colour variations $\sigma_{E(B-V)}
\sim 0.05-0.1$.  

Thermal emission at IR wavelengths could be the ``smoking gun'' for
the presence of CS dust. Pre-existing CS dust may be radiatively
heated by absorption of UV/optical photons from the SN or
collisionally heated by the SN shock. New dust grains could also be
formed in SN Ia ejecta.
\citet{nozawa2011} model this process, and find that up to 0.2~$M_{\odot}$ 
of dust could condense $\sim 100-300$ days after the explosion. 

\citet{gerardy2007} observed two normal SNe Ia (SNe 2003hv and 2005df)
at late phases ($\sim 100-400$ days after explosion) with the {\it
  Spitzer Space Telescope} in the $3.6-22 \mu$m wavelength range. The
mid-IR spectral energy distributions (SEDs) and photometry are
compatible with strong atomic line emission from the SN, and therefore
exhibit no compelling indication of pre-existing or newly formed dust.
\citet{nozawa2011} compare their models with the \citet{gerardy2007}
photometry and derive an upper limit of $0.075\,\mathrm{M}_{\sun}$ of
newly formed silicate dust.
Furthermore, \citet{gomez2012} studied the Kepler and Tycho supernova
remnants (thought to be remnants of SNe Ia that exploded $\sim 400$
years ago) using observations in the $24-850\,\mu$m range and reported
the detection of $\sim 3-9 \times 10^{-3}\,\mathrm{M}_{\sun}$ of warm
dust ($\sim$90~K). Their findings are consistent with the warm dust
originating in the circumstellar (Kepler) and interstellar (Tycho)
material swept up by the primary blast wave of the remnant.

In this paper, we present the earliest far-IR measurements of SNe Ia, within 45 days after explosion, using the {\it Herschel Space Observatory} \citep{pilbratt2010} from 70-160 $\mu$m. We also derive limits on pre-existing dust in the CS environment of
the three observed SNe.

\section{Targets and Observations}
Thermal emission from heated pre-existing CS dust would be difficult
to detect in the near-IR, except for large masses and high
temperatures, but could be detected at mid-IR wavelengths, e.g. with
{\it Spitzer}. However, the degeneracy with the photospheric emission
around $5\, \mu$m makes it challenging to discriminate between emission by
dust and intrinsic light from the SN.  Conversely, observations at
longer wavelengths (beyond $10\,\mu$m) would be dominated by radiating dust,
which motivates the use of {\it Herschel} observations for this study.
\footnote{At the time our targets were observed, only the {\it Spitzer} $3.6$ and $4.5\,\mu$m bands were operational.}

To investigate the presence of CS dust shells, the observations were carried out within 45 days from the SN explosion in order to minimise the risk of confusion with any newly formed dust produced in the SN ejecta \citep[as seen in core-collapse SNe; ][]{kotak2009}. 

Another factor is the duration of the IR echo, which is expected to
scale with the radius of the CS dust shell, $t_{\rm echo} \sim 2r_{\rm
  d}/c$.  For a geometrically thin, spherically symmetric shell, the
fraction of emitting dust mass perceived by the observer increases
with time, reaching maximum at $t=t_{\rm echo}$.

For shell radii, $r_{\rm d} \sim 10^{16}\, \mathrm{cm}$, the IR
echo would be too short to be captured by our observations. However,
in such a scenario, the CS dust would have been heated to high enough
temperatures for its near-IR emission to dramatically change the early
part of the observed lightcurves. 
For dust at radii $r_{\rm d} \sim 10^{17}\,\mathrm{cm}$, the IR echo ($t_{\rm echo} = 3-4$ months) 
is partially within our observing window. 
Thus, although our observational strategy may miss the IR echo
maximum, it nonetheless represents a reasonable compromise for
exploring possible pre-existing CS dust shells. 

In this study we targeted three Type Ia SNe: SNe 2011by, 2011fe and 2012cg, 
selected based on their close proximity. Only one of these three SNe showed 
significant reddening at optical wavelengths (SN 2012cg), thus making a detection at far-IR 
wavelengths more challenging for the remaining two.

\subsection{Herschel PACS data}\label{sec:pacsdata}
The observations of SNe 2011by, 2011fe and 2012cg were obtained using
the {\it Photodetector Array Camera and Spectrometer} \citep[PACS;][]{poglitsch2010} on board
{\it Herschel}.  The mini
scan-map observing mode was used with a scan speed of $20\arcsec$/s,
resulting in a final map of $3\arcmin \times 7\arcmin$ with a homogeneous coverage
in the central region (about $50''$ in diameter).  The full width at
half maximum (FWHM) of the point spread function (PSF) at $70\,\mu$m
and $160\, \mu$m are $6\arcsec$ and $12\arcsec$ respectively.  The flux
calibration uncertainty for the PACS $70$ and $160\,\mu$m bands are currently 
estimated to be smaller than 5\% \footnote{PACS Observer's Manual, Section 3.3.}.
The colour corrections to the modeled dust emission spectra for the 
PACS $70$ and $160\,\mu$m bands are negligible (maximally $\sim 5$\%).

The data reduction was performed up to level2 using the {\it Herschel
  Interactive Processing Environment} \citep[HIPE; ][]{ott2010}.  Each target was
observed for 4 hours, with simultaneous imaging in the $70\, \mu$m and
$160\, \mu$m bands resulting in unconfused $5\sigma$ point source flux limits 
of approximately 5 and 10 mJy respectively.

In addition to our observations we also include analysis of archival
data of the host galaxies of SNe 2011fe \citep[M101 observed as part of the KINGFISH survey;][]{kennicutt2011} 
and 2012cg \citep[NGC 4424 observed as part of the HeVICS survey;][]{davies2012}.  
These observations were carried out in the
large scan mode, with a medium scan map rate of $20\arcsec$/s resulting in
maps with a $3.2\arcsec$/pixel resolution and an unconfused $5\sigma$ point source flux
limit of approximately 25 mJy in the $70\, \mu$m band.

Photometry was performed by using a set of single apertures (with
radii defined by the FWHM of the PSF) to estimate the far-IR flux at the SN positions; apertures were also used to determine the average sky background level on the map, and the background fluxes in the vicinity of the SN.

\begin{figure}
\center\includegraphics[height=0.90\textheight]{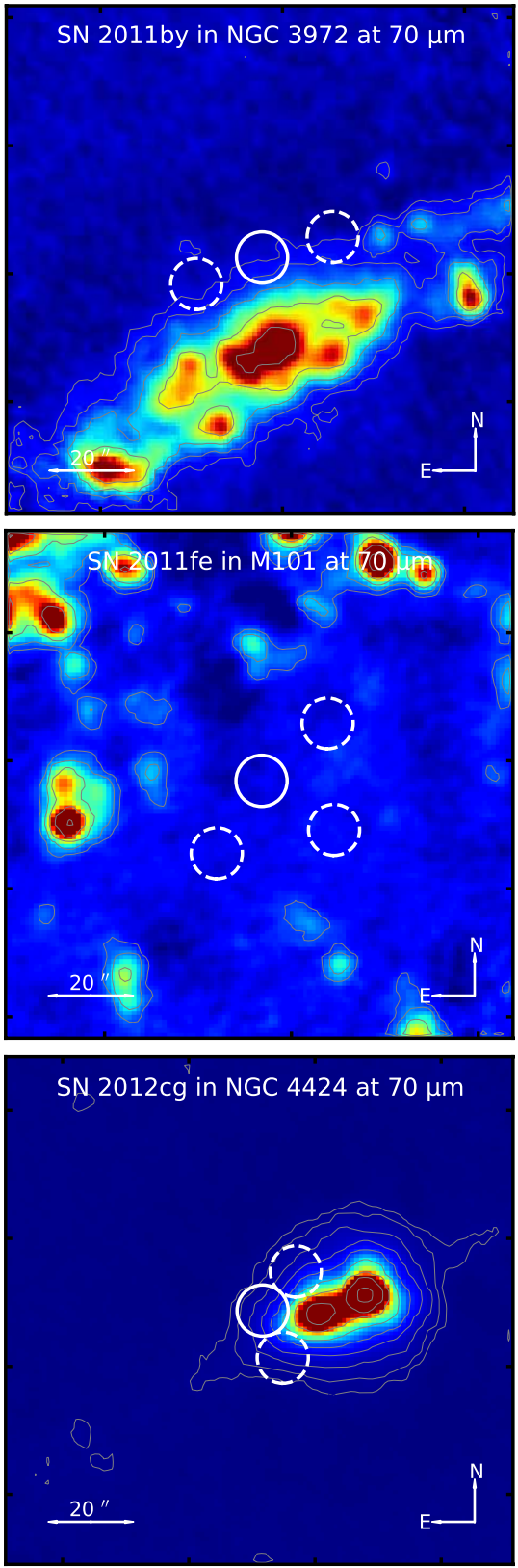}
\caption[]{\label{fig:panels} {\it Herschel} PACS 70$\mu$m observations of
  SNe 2011by (top panel), 2011fe (middle panel) and 2012cg (bottom
  panel). The solid circles indicate the position of the supernovae and the
  FWHM of the PSF ($6\arcsec$). 
  The dashed circles show the apertures used for background estimation.} 
\end{figure}

\subsection{SN 2011by}\label{sec:by}
SN 2011by was discovered 2011/04/26.823 by Zhangwei Jin and Xing Gao
at R.A. = 11:55:45.56, Decl. = +55:19:33.8 at a location $5\arcsec.3$ East and
$19\arcsec.1$ North of the center of the barred spiral galaxy NGC 3972 
\citep[$D=18.5 \pm 0.8\,\mathrm{Mpc}$, ][]{tully2009}.  
The SN reached a peak $B$-band magnitude of $\sim 13$ on 
around May 9 with a colour excess of $E(B-V) \approx 0.08$ mag \citep{maguire2012}. 

Our PACS $70\, \mu$m observations from 2011 May 24 (about two weeks after $B$-band
maximum) are shown in Fig.~\ref{fig:panels}.  
The SN exploded in a region of significant host galaxy background emission.  
To derive upper limits on possible emission from pre-existing CS dust, we
compare the flux at the SN position with the estimated host galaxy
background flux in the vicinity (Tab.~\ref{tab:photometry}).
The galactic emission was estimated by placing apertures along
iso-flux contours.  We measure no significant excess far-IR emission
($3.7 \pm 1.5 \, \mathrm{mJy}$ at $70\, \mu$m) with respect to the 
estimated host galaxy background at the location of SN 2011by .

\subsection{SN 2011fe}\label{sec:fe}
SN 2011fe was discovered 2011/08/24.000 by the Palomar Transient
Factory at R.A. = 14:03:05.81, Decl. = +54:16:25.4 (J2000) at a location
$58\arcsec.6$ West and $270\arcsec.7$ South of the center of the nearby spiral
galaxy M101 \citep[$D=6.4 \pm 0.5 \,
\mathrm{Mpc}$, ][]{shappee2011}.  The SN reached a peak $B$-band
magnitude of $\sim 10$ on around September 10 \citep{matheson2012}.
The Galactic and host galaxy reddening, deduced from the integrated
equivalent widths of the Na I D lines are $E(B-V)_{\rm MW} = 0.011 \pm 0.002$ and
$E(B-V)_{\rm host} = 0.014 \pm 0.002$ mag, respectively
\citep{patat2011}.

By analysing pre-explosion {\it HST} and {\it Spitzer} images,
\citet{li2011} and \citet{nugent2011} are able to 
rule out red-giants and a
majority of helium stars as the mass donating companion to the
exploding WD. Early phase radio and X-ray observations
\citep{horesh2012, chomiuk2012, margutti2012} report non-detections,
yielding constraints on the pre-explosion mass-loss rate from the
progenitor system 
$\dot{M} \lsim 6 \times 10^{-10} - 10^{-8} (v_{\mathrm{wind}} / 100 \, \mathrm{km\,s^{-1}}) \, \mathrm{M}_{\sun}  \mathrm{yr}^{-1}$. 
Although they are model dependent, these limits rule out a large portion of the
parameter space of single-degenerate progenitor models for SN 2011fe.
The absence of time-variant, blueshifted absorption features also rules out the presence
of substantial amounts of CSM \citep{patat2011}. In summary, previous observations are
consistent with the progenitor of SN 2011fe being a binary system with
a main sequence or a degenerate companion star.

Our {\it Herschel} PACS $70\, \mu$m data from 2011
October 02 (about 33 days after $B$-band maximum) are shown in
Fig.~\ref{fig:panels}. SN 2011fe is located in a region with low host
galaxy background emission. No excess far-IR emission is detected at
the position of SN 2011fe ($-1.5 \pm 1.5\, \mathrm{mJy}$ at $70\,
\mu$m). We also analysed archival data to obtain the far-IR flux before the explosion (described in \S~\ref{sec:pacsdata}, Tab.~\ref{tab:photometry}). The measured background subtracted flux at the SN
position is $-4.7 \pm 6.2\, \mathrm{mJy}$.  There is no significant far-IR
source evident at the location of the SN before or after the explosion.

\subsection{SN 2012cg}\label{sec:cg}
SN 2012cg was discovered 2012/05/15.790 by the Lick Observatory
Supernova Search at R.A. = 12:27:12.83, Decl. = +09:25:13.2 (J2000) at a
location $17\arcsec.3$ East and $1\arcsec.5$ South of the peculiar SBa galaxy NGC
4424 \citep[$D=15.2 \pm 1.9 \,\mathrm{Mpc}$, ][]{cortes2008}.  
SN~2012cg reached a peak $B$-band magnitude of $12.1$ on 2012 June 2. The
SN show signs of host galaxy reddening, with a colour excess of $E(B-V)
\approx 0.2$ mag derived from both optical photometry and
high-resolution spectroscopy \citep{silverman2012, ATel4159}.

Our {\it Herschel} PACS $70\, \mu$m data from 2012 June 11
(about 9 days after $B$-band maximum) are shown in
Fig.~\ref{fig:panels}.  SN~2012cg is located in a region of
significant host galaxy far-IR emission.  We derive upper limits on
possible emission from pre-existing CS dust
(Tab.~\ref{tab:photometry}) in a similar manner to SN 2011by, by
comparing the flux at the SN position with the host galaxy background
flux in the vicinity. We measure no excess far-IR emission ($-0.7 \pm 1.8\,\mathrm{mJy}$ at $70\, \mu$m)
with respect to the estimated host galaxy background at the location of SN
2012cg .

In addition, we also analyse pre-explosion archival PACS $70\, \mu$m
data (described in \S~\ref{sec:pacsdata}). The background subtracted flux at the SN location 
is $-15 \pm 5 \,\mathrm{mJy}$. There is no significant far-IR
source evident at the location of the SN before or after the explosion.

\begin{table}
  \centering
  \caption{Photometry of SNe 2011by, 2011fe and 2012cg.}\label{tab:photometry}
  \begin{tabular}{@{}llccc@{}}
    \hline
    Target     	& Host		& Days from	    	& $F_{\nu}^{70\, \mu {\rm m}}$ 	   	& $F_{\nu}^{160\, \mu {\rm m}} $         \\
               	& galaxy	& $B_{\rm max}$	    	& (mJy)	    		& (mJy)	       \\
    \hline\hline    
    SN 2011by  	& NGC 3972	& +15  & $3.7 \pm 1.5$ 	& $16 \pm 8$      \\
    \hline
		& M101$^{\rm a}$	& -451	& $-4.7 \pm 6.2$  	& ---				\\
    SN 2011fe  	& M101		& +23  	& $-1.5 \pm 1.6$ 	& $-16 \pm 10$    \\
    \hline
    		& NGC 4424$^{\rm b}$ & -314	& $-15 \pm 5$  		& ---				\\
    SN 2012cg  	& NGC 4424	& +9	& $-0.7 \pm 1.8$ 	& $-23 \pm 8$     \\
    \hline
  \end{tabular}
  Data from $^{\rm a}$ \citet{kennicutt2011} and $^{\rm b}$ \citet{davies2012}.
\end{table}

\section{Upper limits from dust models}
To model the far-IR emission from pre-existing CS dust we
consider the idealized case \citep[described
in][]{hildebrand1983,fox2010} of an optically thin dust cloud of mass
$M_{\rm d}$ with dust particles of radius $a$, emitting thermally at a
single equilibrium temperature $T_{\rm d}$. The expected flux at a
distance $D$ is,
\begin{equation}
  F_{\nu} = M_{d} \frac{\kappa_{\nu}(a) B_{\nu}(T_{\rm d})}{D^{2}} ,
\end{equation}
where $B_{\nu}(T_{\rm d})$ is the Planck blackbody function and the
dust mass emissivity coefficient, $\kappa_{\nu}(a)$, is
\begin{equation}
\kappa_{\nu}(a) = \left( \frac{3}{4{\rm \pi} \rho a^{3}} \right) {\rm \pi} a^{2} Q_{\nu}(a) = \frac{3 Q_{\nu}(a)}{4 a \rho}.
\end{equation}
$Q_{\nu}(a)$ is the absorption efficiency and the dust bulk (volume)
density, $\rho \approx 2-3\,\mathrm{g/cm}^3$ depending on grain composition.
The expected emission depends on the choice of dust grain composition
and size.  Interstellar dust is well described by a mixture of
silicate and graphitic grains of different sizes, and generally in the
far-IR $\kappa \propto \lambda^{-\beta}$ with $\beta \sim 1-2$ and
$\kappa \approx 67.0\,\mathrm{cm}^2/\mathrm{g}$ at $70\, \mu$m \citep{draineli2001}.
However, CS dust around SNe may well be dominated by either silicate
or graphitic grains depending on the stellar atmosphere of the
involved stars.  Since we do not know the nature of the SNe Ia
progenitor systems and their potential dust production mechanisms, we
will consider separate scenarios of either silicate or graphite grains
of radius $a=0.1\, \mu$m \citep[described
in][]{drainelee1984,laordraine1993,weingartnerdraine2001}.

From the non-detections of the SNe in the PACS $70\, \mu$m and $160\,
\mu$m passbands we calculate upper limits on the CS dust mass
surrounding SNe 2011by, 2011fe and 2012cg.  

\begin{figure}
  \includegraphics[width=84mm]{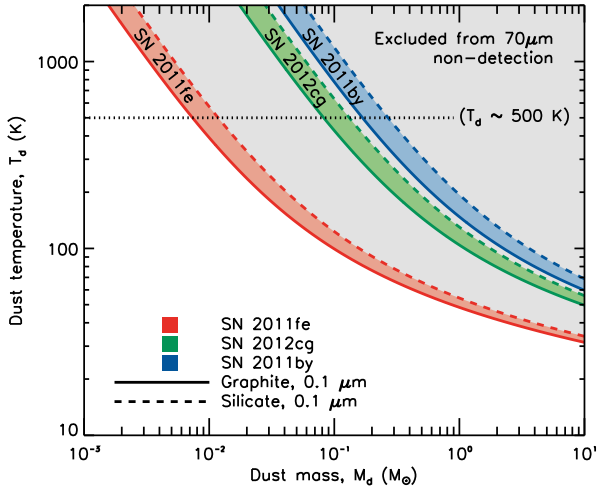}
  \caption{$3\sigma$ upper limits on the circumstellar dust mass and
    temperature, assuming graphitic (solid lines) or silicate (dashed
    lines) dust grains of size $a = 0.1\, \mu$m, derived from the
    non-detection in the {\it Herschel} PACS $70\,\mu$m observations of SNe
    2011by (blue lines), 2011fe (red lines) and 2012cg (green lines).
    The horizontal dotted line indicates the expected temperature
    $T_{\rm d} \sim 500\, \mathrm{K}$ for a pre-existing dust shell of radius
    $r_{\rm d} \sim 10^{17}\, \mathrm{cm}$. }
  \label{fig:contours}
\end{figure}

Fig.~\ref{fig:contours} shows the excluded dust mass range as a function
of temperature for the three SNe, irrespective of heating mechanism.
The upper limit on the dust temperature, set by the evaporation
temperature of the dust grains ($T \lsim 2000\, \mathrm{K}$),
corresponds to a minimal dust survival radius $r_{\rm evap} \sim
10^{16}\, \mathrm{cm}$ \citep{rahman2011}.  
Detections of CSM around SNe Ia have been claimed at somewhat larger
distances, $r_{\rm CSM} \sim 10^{17}\,\mathrm{cm}$
\citep[e.g.][]{patat2007}. To derive an estimate of the expected temperature of CS dust at
similar radii, $r_{\rm d} \sim 10^{17}\,\mathrm{cm}$, we follow the
simple IR echo model in \citet{fox2010} (see their Fig. 8b).  For a
typical peak SN bolometric luminosity of $\sim 10^{9} \,
\mathrm{L}_{\sun}$, radiatively heating a pre-existing dust shell of
radius $r_{\rm d} \sim 10^{17}\, \mathrm{cm}$, graphitic dust grains of $a =
0.1\, \mu$m will be heated to $T_{\rm d} \sim 500\, \mathrm{K}$
(silicate grains would be heated to even higher temperatures).

\begin{figure}
  \includegraphics[width=84mm]{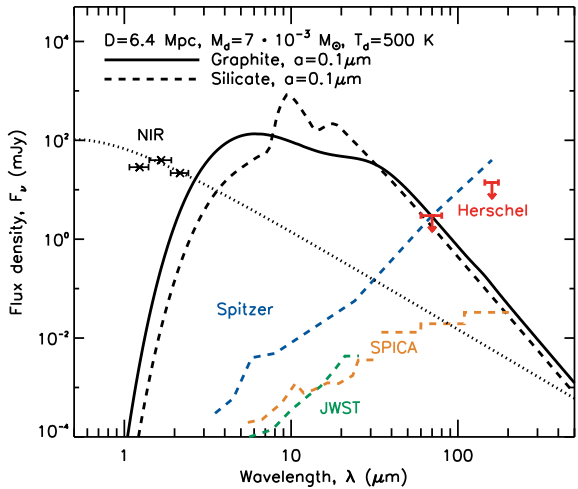}
  \caption{Example of expected IR SEDs of circumstellar dust, assuming
    a distance of 6.4 Mpc, $M_{\rm d}= 7 \times 10^{-3}\,\mathrm{M}_{\odot}$ and
    $T_{\rm d} = 500$ K with graphitic (solid black line) or silicate
    (dashed black line) dust grains of size $a=0.1\, \mu$m.  The dotted
    black line shows a blackbody spectrum at $10^{4}\,\mathrm{K}$, scaled to match
    the NIR fluxes of SN 2011fe 33 days after maximum brightness
    \citep{matheson2012}.  For comparison, the $5\sigma$ detection
    limits of 3 hr observations with {\it Spitzer}
    (blue dashed line), the {\it James Webb Space Telescope} (green
    dashed line) and {\it SPICA} (orange dashed line) are included.  The red symbols indicate the $3\sigma$
    upper limits on the flux of SN 2011fe in the PACS $70\, \mu$m and
    $160\, \mu$m bands (described in \S~\ref{sec:fe}).  }
  \label{fig:dustsed}
\end{figure}

In what follows, we use $T_{\rm d} \sim 500\, \mathrm{K}$ as a point
of reference (marked by the dotted line in
Fig.~\ref{fig:contours}). The expected dust SED for this specific
temperature is shown in Fig.~\ref{fig:dustsed}, along with the
sensitivity of current and future mid- and far-IR facilities.

Due to its proximity, SN 2011fe yields the tightest constraints,
$M_{\rm d} \lsim 7 \times 10^{-3}\,\mathrm{M}_{\odot}$ at a
$3\sigma$-level, assuming graphitic dust grains of size $a=0.1\, \mu$m
heated to temperatures $T_{\rm d} \sim 500\,\mathrm{K}$ (red solid
line in Fig.~\ref{fig:contours}).  For silicate dust grains, the
corresponding upper limit is $M_{\rm d} \lsim 10^{-2}$ M$_{\odot}$
(red dashed line in Fig.~\ref{fig:contours}).  The upper limits for SN
2011by are weaker, $M_{\rm d} \lsim 10^{-1}\, \mathrm{M}_{\odot}$ at a
$3\sigma$-level for simliar assumptions (blue solid and dashed lines
for graphitic and silicate dust grains in Fig.~\ref{fig:contours}).
For SN 2012cg, the upper limits are $M_{\rm d} \lsim 8 \times 10^{-2}\,
\mathrm{M}_{\odot}$ at a $3\sigma$-level for assuming graphitic dust
grains of size $a=0.1\, \mu$m heated to temperatures $T_{\rm d} \sim
500\, \mathrm{K}$ (green solid line in Fig.~\ref{fig:contours}).

\section{Summary and conclusions}
Searches for evidence of CSM around SNe Ia are an important aspect in
the efforts to understand the exact nature of these explosions and
their use as accurate distance estimators. For the latter, the
presence of pre-explosion CS dust could explain the empirically
derived, non-standard reddening corrections that are applied to
minimize the scatter in the SNe Ia Hubble diagram \citep{goobar2008}.

In this work, we searched for far-IR emission from pre-existing CS
dust around three nearby Type Ia SNe {\it within} a few weeks after maximum
brightness.  By considering the {\it Herschel} non-detections, we can
exclude dust masses $M_{\rm d} \gsim 7 \times 10^{-3}\,\mathrm{M}_{\odot}$ for
dust temperatures $T_{\rm d} \sim 500\,\mathrm{K}$ at a $3\sigma$-level for SN
2011fe, and the upper limits are one order of magnitude weaker for SNe
2011by and 2012cg, excluding dust masses $M_{\rm d} \gsim 10^{-1}\,\mathrm{M}_{\odot}$.

Although these are the strictest upper limits on CS dust around newly
exploded SNe Ia, our limits can not completely rule out the presence
of CS dust as a contributing source to SN Ia reddening. Our sensitivity 
for CS dust masses ($M_{\rm d} \sim 10^{-3}$ to
$10^{-2}\,\mathrm{M}_{\odot}$) is about one-two order of magnitudes larger than
the dust masses that have been suggested in simulations \citep[$M_{\rm d}
\sim 10^{-4}\,\mathrm{M}_{\odot}$ in][]{rahman2011}.

While current instrumentation allows mainly for exploration of CS dust
around SNe within the very local universe ($D \lsim 5
\,\mathrm{Mpc}$), future missions such as {\it JWST} and {\it SPICA}, 
will have the potential to dramatically improve the
sensitivity, as shown in Fig.~\ref{fig:dustsed}.

The authors are grateful to the anonymous referee for useful comments
that have significantly improved the paper. We also thank Hugh
Dickinson, Ori Fox and Edvard M\"ortsell for useful discussions and
comments.  RA would like to thank the Swedish National Space Board for
financial support.  AG acknowledges support from the Swedish National
Research Council.

\bibliography{cspaper}

\label{lastpage}
\end{document}